\documentclass[12pt]{article}
\usepackage{amsfonts,amsmath,amssymb}
\usepackage{amsthm,amscd}
\usepackage{subeqnar}
\newfont{\rsfs}{rsfs10 scaled \magstep1}
\newcommand{\scr}[1]{\mbox{\rsfs {#1}}}
\makeatletter

\@addtoreset{equation}{section}
\makeatother
\theoremstyle{definition}
\newtheorem{thm}{Theorem}[section]

\newtheorem{pr}[thm]{Proposition}
\newtheorem{rem}[thm]{Remark}

\newcommand{\pl}{\partial}

\newcommand{\bib}[2]{\frac{\partial {#1}}{\partial {#2}}}
\newcommand{\obib}[2]{\frac{d {#1}}{d {#2}}}

\begin{document}
\title{Applications of the Ashtekar gravity \\
to four dimensional hyperk\"ahler geometry \\
and Yang-Mills instantons}
\author{${}^{\P}$\thanks{E-mail address 
: hashimot@sci.osaka-cu.ac.jp} Yoshitake HASHIMOTO ,
\thanks{E-mail address : yasui@sci.osaka-cu.ac.jp  } Yukinori YASUI , \\
\thanks{E-mail address : miyagi@sci.osaka-cu.ac.jp} Sayuri MIYAGI \   and
\thanks{E-mail address : ootsuka@sci.osaka-cu.ac.jp} Takayoshi OTSUKA  \\
{\small ${}^{\P} $Department of Mathematics, Osaka City University,
Sumiyoshiku, Osaka, Japan} \\
{\small Department of Physics, Osaka City University, 
Sumiyoshiku, Osaka, Japan} }
\date{}
\maketitle


\begin{abstract}
The Ashtekar-Mason-Newman equations are used to construct
the hyperk\"ahler metrics on four dimensional manifolds.
These equations are closely related to
anti self-dual Yang-Mills equations of the infinite dimensional
gauge Lie algebras of all volume preserving vector fields.
Several examples of hyperk\"ahler metrics are presented through
the reductions of anti self-dual connections.
For any gauge group anti self-dual connections on hyperk\"ahler
manifolds are constructed using the solutions of both
Nahm and Laplace equations.

\end{abstract}

\newpage
\section{Introduction}
Four dimensional hyperk\"ahler geometry has been studied extensively
in connection with several issues: gravitational instantons,
supersymmetric non-linear $\sigma$ models and compactifications
of superstrings~\cite{E-G-H,string1,string2}. 
A hyperk\"ahler manifold is a Riemannian manifold equipped with
three covariantly constant complex structures $J^a \, (a=1,2,3)$
which obey the condition 
$J^aJ^b=-\delta_{ab}-{\epsilon_{abc}}J^c$.
In four dimension, a Riemannian manifold is hyperk\"ahler if and
only if its Ricci curvature is zero and its Weyl curvature is
either self-dual or anti self-dual, namely, 
the metric is the solution of
(anti) self-dual Einstein equation.

It is known that all asymptotically locally Euclidean (ALE)
hyperk\"ahler manifolds are classified by Dynkin diagrams of ADE types.
These spaces are constructed as hyperk\"ahler quotients
of flat Euclidean spaces~\cite{Kro}.

On the other hand, in the Hamiltonian approach for general relativity,
Ashtekar and Mason-Newman reduced the problem of finding hyperk\"ahler
metrics to that of finding linearly independent four vector fields 
$V_{\mu} \, (\mu=0,1,2,3)$
and a volume form $\omega$  on a four dimensional manifold $X$
that satisfy the following two conditions~\cite{Ash,M-N,A-J-S}:
\begin{description}
\item[(1)] volume preserving condition
  \begin{equation}\label{volpre}
    L_{V_{\mu}}\omega = 0
  \end{equation}
\item[(2)] half-flat condition
  \begin{eqnarray}\label{nahm}
    \frac12 {\bar{\eta}^a}_{\mu \nu} [ V_{\mu} , V_{\nu}] =0,
  \end{eqnarray}
where ${{\bar{\eta}}^a}_{\mu \nu} \, (a=1,2,3)$ are the 't Hooft
matrices satisfying the relations:
\begin{equation}\label{'tHooft}
  {\bar{\eta}^a}_{\mu \nu} = -{\bar{\eta}^a}_{\nu \mu},
  \qquad {\bar{\eta}^a}_{\mu \nu}{\bar{\eta}^b}_{\mu \sigma}
  = {\delta}_{ab} {\delta}_{\nu \sigma} + {\epsilon}_{abc}
  {\bar{\eta}^c}_{\nu \sigma}.
\end{equation}
\end{description}
The hyperk\"ahler metric on $X$ is given by 
$g \, (V_{\mu},V_{\nu})=\phi \delta_{\mu\nu}$
with $\phi=\omega(V_0,V_1,V_2,V_3)$ (we choose the sign
of $\omega$ such that $\phi$ is positive.).
Then three complex structures $J^a$ are expressed by
\begin{equation}
  J^{a} (V_{\mu}) = {\bar{\eta}^a}_{\nu \mu} V_{\nu}.
\end{equation}

Conversely the pair ($V_{\mu},\omega$) can be locally constructed
from any hyperk\"ahler structure $(g, J^{a} )$ as 
follows~\cite{M-N,Don1}:
for a harmonic function $\tau$ and the volume form $\omega_{g}$
with respect to $g$, the vector fields $V_{\mu}$ are defined by
\begin{subeqnarray}
  V_{0} &=& \phi \, \mbox{grad} \, \tau,\\
  V_{a} &=& - J^{a} (V_0), \, (a=1,2,3),
\end{subeqnarray}
where $\phi = g \, (\mbox{grad} \tau , \mbox{grad} \tau )^{-1}$.
Then $V_{\mu}$ preserve the volume form 
$\omega = {\phi}^{-1} {\omega_{g}}$ and satisfy the half-flat 
condition.
We can choose the local coordinates such that 
$V_0 = \frac{\partial}{\partial \tau}$, and 
then (\ref{nahm}) reduces to the Nahm equation,
\begin{equation}
  \label{Vnahm}
  \frac{\partial V_a}{\partial \tau} =\frac12 \epsilon_{abc} [V_b,V_c],
\end{equation}
where $V_{a} \, (a=1,2,3)$ are volume preserving vector fields on a
$3$-ball.
This form was given by Ashtekar, Jacobson and
Smolin~\cite{Ash,A-J-S}.

In this paper we present a explicit construction of hyperk\"ahler
metrics based on (\ref{volpre}) and (\ref{nahm}).
We also give a new construction of anti self-dual
Yang-Mills connections on
hyperk\"ahler manifolds generalizing the multi-instanton
Ansatz of 't Hooft, Jackiw-Nohl-Rebbi~\cite{J-N-R}
and further Popov~\cite{Pop1,G-M,BFRM}.


\section{Reductions of the Ashtekar-Mason-Newman~equations}
We start with the construction of four vector fields $V_{\mu}$
corresponding to Gibbons-Hawking metrics.
We use the standard coordinates $(x^0,x^1,x^2,x^3)$ and the volume
form $\omega=dx^0\wedge dx^1\wedge dx^2\wedge dx^3$ for the underlying
space-time ${\Bbb R}^4$.
Let $\phi$ and $\psi_i \, (i=1,2,3)$ be smooth functions 
and define~\cite{Joy} 
\begin{subeqnarray}
  \label{vector}
  V_0 &=& \phi \bib{}{x^0}, \\
  V_i &=& \bib{}{x^i} + \psi_i \bib{}{x^0}.
\end{subeqnarray}
Then the volume preserving condition implies that the functions 
$\phi,\, \psi_i$ are independent of $x^0$ and the half-flat condition
implies 
\begin{equation}
  \ast d \phi = d \psi,
\end{equation}
where $\psi = \sum_{i=1}^3 \psi_i dx^i$ and $\ast$ denotes the Hodge
operator on the three dimensional subspace 
${\Bbb R}^3 =\{(x^1,x^2,x^3)\}$.
These conditions are identical to the Ansatz used by Gibbons-Hawking
to construct hyperk\"ahler metrics with a triholomorphic U(1)
symmetry~\cite{G-H}.

\rem
The Gibbons-Hawking Ansatz is characterized by the following Lie 
algebra ${\frak g}_{GH}$. 
Let ${\frak g}_{GH}$ be the Lie algebra generated by the
Gibbons-Hawking vector fields $V_{\mu}$.
Then ${\frak g}_{GH}$ is given by the extension of the two abelian Lie 
algebras $\left< \pl_I \phi \bib{}{x^0} \right>$ and ${\Bbb R}^3$.
($\pl_I \phi$ denotes the multiple partial differentiation of $\phi$ 
with respect to $x^1,x^2 \, \mbox{and}\, x^3$.)
In other words the following sequence is exact,
\begin{equation}
0 \ \to \ \left< \pl_I \phi \bib{}{x^0} \right> \ \to \ {\frak g}_{GH}
\ \to \ {\Bbb R}^3 \ \to \ 0.  
\end{equation}
We note that $\left< \pl_I \phi \bib{}{x^0} \right>$ is a left 
${\scr D}$-module for 
\begin{equation}
  {\scr D} = \left.
  {\Bbb R}\left[ \bib{}{x^1},\bib{}{x^2},\bib{}{x^3}\right]
  \right/ \left( \left( \bib{}{x^1} \right)^2 +
   \left( \bib{}{x^2} \right)^2 + \left( \bib{}{x^3} \right)^2
  \right).
\end{equation}

\quad \\

We now describe an approach which allows us to
obtain the vector fields $V_{\mu}$ satisfying the conditions
(\ref{volpre}) and (\ref{nahm}) from (anti) self-dual 
 Yang-Mills connections of infinite dimensional
gauge groups. Let $\Sigma^{(n)}\, (n=1,2,3,4)$ be
a $n$ dimensional manifold equipped with a volume form 
$\omega^{(n)}$. Then we assume that the gauge Lie algebra
is the Lie algebra $sdiff (\Sigma^{(n)}) $ of all volume
preserving vector fields on $\Sigma^{(n)}$.
Connections on Euclidean space 
${\Bbb R}^4 = \{ (x^0,x^1,x^2,x^3)\}$ may be explicitly 
expressed by $1$-forms on ${\Bbb R}^4$ valued in 
$sdiff(\Sigma^{(n)})$. We write these $1$-forms as
$A_{\mu} dx^{\mu}\, (\mu = 0,1,2,3)$ and require the 
following conditions:
\begin{description}
\item[(1)]
$A_{\mu}$ are ${\Bbb R}^n$-invariant with respect to the coordinates
$(x^0, \cdots , x^{n-1})$.
\item[(2)]
$A_{\mu}$ are anti self-dual connections, namely, 
the covariant differentiations
$ D_{\mu}=\frac{\partial}{\partial x^{\mu}} + A_{\mu}$
satisfy the equation
\begin{equation} 
\frac12 {\bar{\eta}^a}_{\mu \nu} [ D_{\mu}, D_{\nu} ]=0. \label{ASDYMeq.}
\end{equation}
\item[(3)]
  $A_{\mu} \, (0 \le \mu \le n-1) $ are  linearly independent
at each point on $\Sigma^{(n)}$.
\end{description}

We then define the four vector fields $V_{\mu}$ on 
$\Sigma^{(n)} \times {\Bbb R}^{4-n}$ as follows:
\begin{eqnarray}
  \label{D-V}
V_{\mu}=
\left\{  
  \begin{array}{lr}
  {A_{\mu}} & (0 \leq \mu \leq n-1 ),\\
  {D_{\mu}} & (n \leq \mu \leq 3 ).
  \end{array}
\right.
\end{eqnarray}
These vector fields evidently preserve the volume form 
$\omega = \omega^{(n)} \wedge d x^{n} \wedge \cdots \wedge d x^3$
and satisfy the half-flat condition (\ref{nahm}) and hence induce
a hyperk\"ahler metric on $ \Sigma^{(n)}\times {\Bbb R}^{4-n} $.

We note the previous Gibbons-Hawking vector fields can be
obtained by applying the above construction to the case 
$\Sigma^{(1)}={\Bbb R}$.
( In the cases of $\Sigma^{(4)}$ and $\Sigma^{(3)}$, we simply recover
the equations
(\ref{volpre}), (\ref{nahm}) and (\ref{Vnahm}).)

We now concentrate our attention on the explicit construction
of hyperk\"ahler metrics in the case of 
$X= \Sigma \times {\Bbb R}^2 $, where $( \Sigma, \omega_{\Sigma} )$  
is a two dimensional symplectic manifold.
Let us assume that the ${\Bbb R}^2$-invariant connections
$A_{\mu}$ are the Hamiltonian vector fields $X_{f_{\mu}}$ along
$\Sigma \times \{(x^2,x^3)\} $ associated with some functions 
$f_{\mu} \, (\mu = 0 ,1,2, 3 )$ on $\Sigma \times {\Bbb R}^2 $.
Then (\ref{D-V}) becomes
\begin{subeqnarray}\label{sympV}
  V_0 &=& X_{f_0}, \\
  V_1 &=& X_{f_1}, \\
  V_2 &=& \bib{}{x^2} + X_{f_2}, \\
  V_3 &=& \bib{}{x^3} + X_{f_3}. 
\end{subeqnarray}
Thus our problem consists in finding the four functions $f_{\mu}$
satisfying the anti self-dual condition (\ref{ASDYMeq.}). 
We will not attempt to answer this question in any generality,
but consider three examples below.

\paragraph{(i)} 
Let $f_{\mu}\, (\mu = 0 ,1, 2)$ be $x^2$-independent
and $f_3=0$. Then (\ref{ASDYMeq.}) yields
the Ward equation~\cite{Ward1}:
\begin{equation}
  \label{fnahm}
 \frac{\partial f_0}{\partial x^3} =\{ f_1 , f_2 \},\, 
 \frac{\partial f_1}{\partial x^3} =\{ f_2 , f_0 \}, \,
 \frac{\partial f_2}{\partial x^3}=\{ f_0 , f_1 \},
\end{equation}
 where $\{ \, , \, \}$ denotes the Poisson bracket on $\Sigma$
induced by the symplectic structure $\omega_{\Sigma}$.
These equations ( but with Poisson brackets replaced by commutators
of matrices of some finite dimensional Lie algebra )
arose in Nahm's construction of monopole solutions in Yang-Mills
theory~\cite{Nahm,A-H}.

The group $SL(2,{\Bbb R})$ acts on $\Sigma= {\Bbb R}^2 , H$
(the complex upper half-plane) and $SU(2)$ acts on $\Sigma=S^{2}$
preserving the standard volume form for each case .
So we can construct hyperk\"ahler metrics on 
$\Sigma \times {\Bbb R} \times I $ ($I$ is an open interval)
from the solutions of the Nahm equation valued in $sl(2, {\Bbb R})$
or $su(2)$.
For example, we can express such solutions by Jacobi elliptic functions 
as follows:
\begin{description}
\item[(a)] $sl(2,{\Bbb R})$
\begin{subeqnarray}
\hspace{-5em}
f_0 &=& k \, \textrm{sn}(x^3,k) h_0, \\
\hspace{-5em}
f_1 &=& k \, \textrm{cn}(x^3,k) h_1, \\
\hspace{-5em}
f_2 &=& \textrm{dn}(x^3,k) h_2,
\end{subeqnarray}
\item[(b)] $su(2)$
\begin{subeqnarray}
\hspace{-5em}
f_0 &=& \textrm{ns}(x^3,k) {\hat{h}}_0, \\
\hspace{-5em}
f_1 &=& \textrm{ds}(x^3,k) {\hat{h}}_1, \\
\hspace{-5em}
f_2 &=& \textrm{cs}(x^3,k) {\hat{h}}_2,
\end{subeqnarray}
\end{description}
\begin{eqnarray*}
\hspace{-7em}
\mbox{where} \quad
&& k \in {\Bbb R} \setminus \{ 0 \} \qquad \mbox{and} \\ 
&& h_0 = \{ h_1 ,h_2 \}, \quad  h_1 = -\{h_2,h_0 \}, 
\quad h_2 = -\{ h_0, h_1 \}, \\ 
&& {\hat{h}}_0 = - \{ {\hat{h}}_1, {\hat{h}}_2 \},
\quad {\hat{h}}_1 =- \{ {\hat{h}}_2,{\hat{h}}_0 \}, \quad 
{\hat{h}}_2 = - \{ {\hat{h}}_0 , {\hat{h}}_1 \}.
\end{eqnarray*}
Some explicit solutions of the Nahm equations valued in finite
dimensional Lie algebras were constructed 
in~\cite{Ward1,Ward2,C-A-C}. 

\paragraph{(ii)}
We consider the hyperk\"ahler metric with one rotational Killing
symmetry preserving one complex structure but not 
triholomorphic~\cite{LeB}.
We take $\Sigma = {\Bbb R}^2$ with the coordinates $(y^0,y^1)$
and introduce a $y^0$-independent function $\psi(y^1,x^2,x^3)$.
Let us assume the form,
\begin{subeqnarray}\label{azToda}
  f_0 &=& - 2 e^{\frac{\psi}{2}} \cos \frac{y^0}{2},\\  
  f_1 &=& 2 e^{\frac{\psi}{2}} \sin \frac{y^0}{2},\\
  f_2 &=& -{\int}^{y^1} 
      \frac{\partial \psi}{\partial x^3} dy^1,\\
  f_3 &=& {\int}^{y^1} 
      \frac{\partial \psi}{\partial x^2} dy^1.
\end{subeqnarray}
Then the functions $f_{\mu}$ satisfy the anti self-dual condition 
(\ref{ASDYMeq.}),
if $\psi$ is a solution of three dimensional continual Toda equation:
\begin{equation}
  {\left( \frac{\pl }{\pl x^2}\right)}^2 \psi + 
{\left( \frac{\pl }{\pl x^3} \right) }^2 \psi 
+ {\left( \frac{\pl }{\pl y^1} \right) }^2 e^{\psi} =0. 
\end{equation}
This solution leads to a hyperk\"ahler metric with the Killing vector field
$\bib{}{y^0}$, which is known as the real heaven solution in the Pleba\'nski
formalism~\cite{B-F}.

\paragraph{(iii)}
Using the solutions of both Nahm and Laplace equations,
Popov presented a construction of self-dual Yang-Mills
connections on ${\Bbb R}^4$~\cite{Pop1}.
We may apply the same method to our case, 
i.e. ${\Bbb R}^2$-invariant Yang-Mills connections of the gauge
Lie algebra $sdiff(\Sigma)$.
Let $T^a(t) \, (a=1,2,3)$ and $u(x^2,x^3)$ be solutions of the Nahm
equation associated with the Lie algebra $sdiff(\Sigma)$
and the Laplace equation on ${\Bbb R}^2$, respectively:
\begin{subeqnarray}
  &&\frac{\partial T^{a}}{\partial t} = 
    \frac12 {\epsilon}_{a b c}
    \{ T^{b} , T^{c} \} \quad (a,b,c=1,2,3),\label{unahm}\\
  && \sum_{\mu=2}^{3} \left(\bib{}{x^{\mu}}\right)^2  u=0.
 \label{Laplace}
\end{subeqnarray}
Then we obtain the solutions $f_{\mu} \,
(\mu=0,1,2,3)$:
\begin{equation}  \label{NLansats}
  f_{\mu} = \sum_{\nu=2}^{3}{\bar{\eta}^a}_{\mu \nu}
  \bib{u}{x^{\nu}} T^a(u).
\end{equation}

\rem
We note that the Ansatz for the vector fields (\ref{sympV}) is similar
to that in~\cite{Gra,Hus,Str}, which should be regarded as the Ansatz
for the vector fields on a complex four dimensional manifold,
so that it is not so obvious whether the corresponding metrics
satisfy the reality condition.
However, our construction manifestly satisfies such a condition.

\section{Yang-Mills instantons on hyperk\"ahler manifolds}

Now we present a construction of anti self-dual Yang-Mills 
connections of arbitrary gauge Lie algebra ${\frak g}$ on a
hyperk\"ahler manifold $X$.
As mentioned above, Popov have obtained a formula for 
self-dual Yang-Mills connections on ${\Bbb R}^4$.
We generalize their construction to the case of 
four dimensional hyperk\"ahler
manifolds by using Ashtekar variables.

Suppose $X$ is a hyperk\"ahler manifold expressed by linearly
independent four vector fields $V_{\mu}$ and a volume form $\omega$
as mentioned in (\ref{volpre}) and (\ref{nahm}).
\begin{pr}
  The ${\frak g}$-valued connection $A$ on $X$ given by
  \begin{equation} \label{ym}
    A(V_{\mu}) = {\bar{\eta}^a}_{\mu\nu}(V_{\nu}u)T^a(u)
  \end{equation}
  satisfies the anti self-dual condition if and only if the 
  set of ${\frak g}$-valued functions
  $T^a \, (a=1,2,3)$ on ${\Bbb R}$ is a solution of the Nahm equation:
  \begin{equation}\label{Tnahm}
    \frac{d {T^a}}{d t} = \frac12 {\epsilon}_{abc}[T^b,T^c],  
  \end{equation}
  and the function
  $u$ on $X$ is a solution of the equation:
  \begin{equation}\label{lap}
    \sum_{\mu=0}^3 (V_{\mu}V_{\mu})u = 0. 
  \end{equation}
\end{pr}
\begin{description}
\item[Proof]
Using the identities (\ref{'tHooft}),
we calculate the anti self-dual part of the curvature
$F_{\mu\nu}=F(V_{\mu},V_{\nu})$ as follows:
  \begin{eqnarray*}
    \frac12 {\bar{\eta}^a}_{\mu\nu}F_{\mu\nu} &=& 
      \frac12 {\bar{\eta}^a}_{\mu\nu}\{V_{\mu}(A(V_{\nu}))-
           V_{\nu}(A(V_{\mu})) - A([V_{\mu},V_{\nu}]) +
             [A(V_{\mu}),A(V_{\nu})] \} \\
      &=& -  \sum_{\sigma}(V_{\sigma}V_{\sigma}u) T^a 
      - \sum_{\sigma}(V_{\sigma}u)(V_{\sigma}u)     
    \left\{ \obib{T^a}{t}-\frac12 \epsilon_{abc}[T^b,T^c]\right\}
       \\
      &&-\frac12 A({\bar{\eta}^a}_{\mu\nu}[V_{\mu},V_{\nu}])
  \end{eqnarray*}
It follows from (\ref{nahm}) and our assumptions (\ref{Tnahm})
and (\ref{lap}) that the above quantity 
vanishes. \qed
\end{description}

\rem
The equation (\ref{lap}) is identical to the Laplace equation
associated with the hyperk\"ahler metric.
So we can choose the harmonic functions as $u$ in (\ref{lap}).
For example, if the solutions of (\ref{nahm}) are the Hamiltonian
vector fields for functions $f_{\mu}$ on a symplectic four dimensional
manifold, then $f_{\mu}$ are harmonic.
\rem
In the above proof we don't use the volume preserving condition for the vector 
fields $V_{\mu}$.
Hence it is not necessary that $X$ is a 
four dimensional hyperk\"ahler manifold.
The restriction for a $n$ dimensional
manifold $X$ we require is simply the existence of 
the linearly independent $n$ vector fields $V_{\mu}$
such that $\frac12 {\bar{\eta}}^a_{\mu\nu}[V_{\mu},V_{\nu}]=0$
with the relations (\ref{'tHooft})
for the matrices $\bar{\eta}^a_{\mu\nu}$\cite{Pop2}.


\end{document}